\documentclass[prd,
               twocolumn,
               amssymb,
               amsmath,
               eqsecnum,
               showpacs,
               letterpaper,
               superscriptaddress,
               altaffilletter]
               {revtex4}

\usepackage{color}
\usepackage{graphicx}
\usepackage{latexsym}
\usepackage{float}
\usepackage{amsmath}
\usepackage{amssymb}
\usepackage{multirow}
\RequirePackage{fix-cm}

\pdfoutput=1

\begin{document}

\title{Regression of Environmental Noise in LIGO Data}

\author{V.~Tiwari}  
\affiliation{University of Florida, P.O.Box 118440, Gainesville, Florida, 32611, USA}
\author{M.~Drago}  
\affiliation{University of Trento, Physics Department and INFN, Trento Institute for Fundamental Physiscs and Applications, 
             via Sommarive 14, 38123 Povo, Trento, Italy}
\affiliation{Max Planck Institut f\"{u}r Gravitationsphysik, Callinstrasse 38, 
             30167 Hannover, and Leibniz Universit\"{a}t Hannover, Hannover, Germany}
\author{V.~Frolov} 
\affiliation{LIGO Livingston Observatory, LA , USA}
\author{S.~Klimenko}  
\affiliation{University of Florida, P.O.Box 118440, Gainesville, Florida, 32611, USA}
\author{G.~Mitselmakher}
\affiliation{University of Florida, P.O.Box 118440, Gainesville, Florida, 32611, USA}
\author{V.~Necula}
\affiliation{University of Florida, P.O.Box 118440, Gainesville, Florida, 32611, USA}
\author{G.~Prodi} 
\affiliation{University of Trento, Physics Department and INFN, Trento Institute for Fundamental Physiscs and Applications, 
             via Sommarive 14, 38123 Povo, Trento, Italy}
\author{V.~Re} 
\affiliation{INFN, Sezione di Roma Tor Vergata, Via della Ricerca Scientifica 1, 00133 Roma, Italy}
\author{F.~Salemi} 
\affiliation{Max Planck Institut f\"{u}r Gravitationsphysik, Callinstrasse 38, 
             30167 Hannover, and Leibniz Universit\"{a}t Hannover, Hannover, Germany}
\author{G.~Vedovato}  
\affiliation{INFN, Sezione di Padova, via Marzolo 8, 35131 Padova, Italy}             
\author{I.~Yakushin} 
\affiliation{LIGO Livingston Observatory, LA , USA}

\begin{abstract}
We address the problem of noise regression in the output of gravitational-wave (GW) interferometers, using data from the physical environmental monitors (PEM). The objective of the 
regression analysis is to predict environmental noise in the gravitational-wave channel from the PEM measurements. One of the most promising regression method is 
based on the construction of Wiener-Kolmogorov filters. Using this method, the seismic noise cancellation from the LIGO GW channel has already been performed. In the 
presented approach the Wiener-Kolmogorov method has been extended, incorporating banks of Wiener filters in the time-frequency domain, multi-channel analysis and regulation schemes, 
which greatly enhance the versatility of the regression analysis. Also we presents the first results on regression of the bi-coherent noise in the LIGO data.
\end{abstract}

\date[\relax]{Dated: \today }

\maketitle


\section{Introduction}

Laser Interferometer Gravitational-Wave Observatory (LIGO) is a large-scale physics experiment 
targeting the first direct detection and study of gravitational waves from astrophysical 
sources. Two LIGO detectors in Livingston, LA and Hanford, WA are being upgraded to increase
their sensitivity by an order of magnitude and plan to start taking data in 2015. 

The LIGO detectors employ a Michaelson type interferometer, which measures the differential 
length change induced  by a passing gravitational wave in its two perpendicular arms with 
the length of 4km each~\cite{LIGO}. While the fundamental sources of noise (thermal, quantum, 
shot, etc) determine the baseline LIGO sensitivity, the actual detector noise may also have 
a significant contribution from the environment. Numerous sources, such as ambient magnetic field, 
power grid mains, acoustic and seismic disturbances are coupled into the detector sub-systems 
and can be observed as spectral lines or relatively broadband humps in the detector output, 
degrading the overall detector sensitivity. 

In addition to the LIGO's GW channel, 
data from thousands of Physical Environmental Monitors (PEM) is collected 
to characterize the coupling between the GW channel and the environment. 
Many environmental disturbances recorded in the detectors are due to the linear coupling to 
the environment, for example, the power grid harmonics (power lines). 
Also the detector data contains the up-conversion  noise, which is produced by the interference 
of two or more environmental noises. Most prominent example of the up-conversion noise is the 
bi-linear coupling of the seismic noise which appears as side-bands around the power lines 
and calibration lines. To improve the data quality and increase the detection sensitivity, 
several methods have been used to identify and ultimately remove the environmental noise from 
the detector data~\cite{Coughlin}. 
The examples include the removal of seismic noise using the array of seismometers~\cite{depa}, removal of the line noise with the finite impulse response (FIR) 
filters~\cite{Searle}, removal of the cross-talk between an auxiliary channel and the channel of interest by using the frequency-domain linear regression~\cite{Allen}, 
and prediction of GW strain data channel using Michelson channel noise~\cite{mkr2014}.


In this paper we address the problem of the noise regression (prediction and cancellation) by constructing a special bank of the  Wiener-Kolmogorov filters. 
 The environmental noise contribution to 
the GW channel (target channel) is predicted by filtering data from one 
or more PEM channels (witness channels) simultaneously. 
To capture the frequency dependent correlation between the target and the witness channels and 
for better estimation of the filter parameters, the regression analysis is
performed in the time-frequency (wavelet) domain.
Introduction of multiple witness channels in predicting the target channel noise increases the effectiveness of the constructed filter banks. 
The strength of the presented regression analysis lies not only in predicting the 
linear, but also the up-conversion environmental noise. It is achieved by
creating the synthetic witness 
channels from PEM and simulated channels, which mimic the physical process of the up-conversion. 
This becomes useful as there are no witness channels that measure bilinear noise 
directly. The proposed regression model is expected to be more applicable when a new generation of advanced detectors become operable.

The paper is organized as follows: Section~\ref{regana} discusses the regression analysis; Section~\ref{app} presents some applications of the regression 
analysis; paper is concluded in section~\ref{conclusion}.

\section{Regression Analysis \label{regana}}
In application to the regression 
of LIGO data, we consider the Wiener-Kolmogorov (WK) filters~\cite{Wiener}. Given sampled data from a 
selected auxiliary channel $x$ measuring environmental noise (witness channel), a finite 
impulse response (FIR) filter $a$ can be constructed to predict the noise contribution $n$ 
into the GW channel $t$ (target channel). 
The output of the filter is given by the expression
\begin{equation} 
n[i]=\sum_{j=-L}^{L}{a_{j}x_{i+j}},  
\label{single_channel_reg}
\end{equation}
where $a_j$ are the filter coefficients and $2L+1$ is the filter length.
The filter is obtained by minimizing the mean-square error  
\begin{equation} 
\chi^2=\sum_{i=L}^{N+L}\left(t[i]-\sum_{j=-L}^{L}a_{j}x_{i+j}\right)^2, 
\label{wkhchis}
\end{equation}
where $N$ is the number of data samples used for the estimation (training) of the filter. 
The filter coefficients $a_j$ are calculated by solving the Wiener-Hopf (WH) equations
\begin{equation} R_{xx}\textbf{a}=\textbf{p}_{tx}. \label{wkh}\end{equation}
where $\textbf{p}_{tx}$ is the vector with $2L+1$ components representing the 
cross-correlations between the witness channel with the target channel and $R_{xx}$ is $(2L+1)\times(2L+1)$  auto-correlation matrix for the witness 
channel. The matrix $R_{xx}$ is positive definite and therefore non-singular, 
yielding a unique solution for the filter coefficients. 
Usually, the solution is obtained with the Levinson-Durbin algorithm~\cite{Durbin} 
without the explicit inversion of $R_{xx}$. The predicted noise is calculated by using equation \ref{single_channel_reg} and 
can be subtracted from the target channel, thus reducing the noise in the GW data.


A special case of the Wiener filter is the linear prediction error filter, which is 
obtained by minimizing the following equation
\begin{equation} \chi^2=\sum_{i=L}^{N+L}\left(t[i]-\sum_{j=-L, j\neq 0}^{L}a_{j}t_{i+j}\right)^2. 
\label{lipr}
\end{equation}
In this case, the data sample $i=L$ is predicted by the surrounding samples in the same time series.
 This is well suited for the prediction of a quasi-monochromatic noise in the target channel.

The RMS value of the witness channel data can be broken down into coupled and uncoupled parts. The target and witness channel 
data are whitened before being presented to the analysis, e.g., has a RMS value of 1. Following this, one can write,
\begin{equation} 1 = r_c^2+r_u^2, \end{equation}
where $r_c/r_u$ is the RMS of the coupled/uncoupled part in the witness data.

\subsection{Regression with WDM}
\begin{figure}[htbp]
 \begin{center} 
 \includegraphics[width=0.5\textwidth]{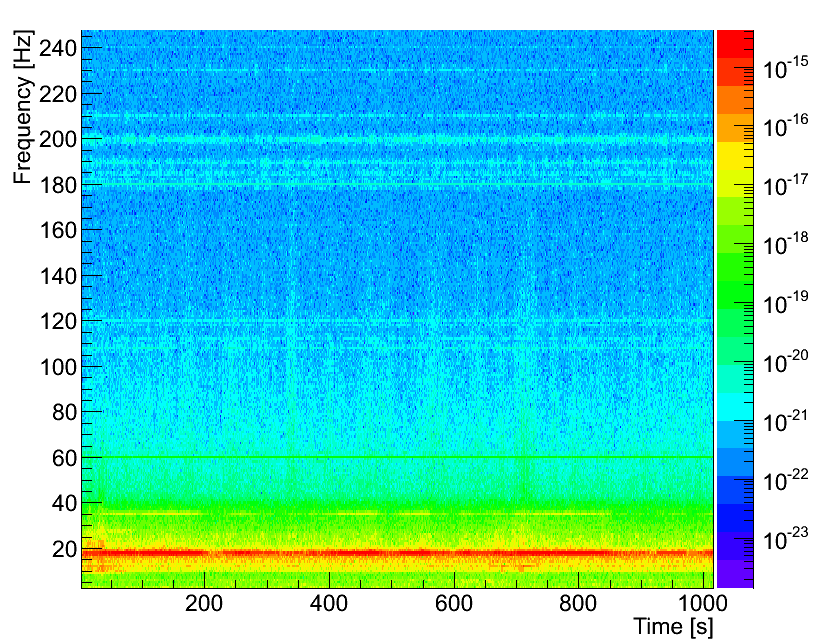}
 \end{center}
\caption{\label{WDM} An example of time-frequency map obtained by applying WDM transform on LIGO data (not all layers have been shown for clarity purposes). Layers have a 
bandwidth of 1 Hz. Regression analysis is performed for each layer separately. Once the prediction is calculated and subtracted from each frequency sub-band 
of the target data, the inverse transform is applied to bring the data back to the time domain.}
\end{figure}

The interferometer data span a wide frequency band (0-8 kHz) and have 
a large dynamic range. Therefore, construction of the WK filters in the time domain requires 
long filters. For example, to regress power line harmonics (15 in the 0-1 kHz band) a WK 
filter with approximately 10000 coefficients should be constructed. Apart from 
the computational complexity associated with the inversion of the $10000{\times}10000$ matrix, 
one need to determine accurately $10000$ filter coefficients, while only 30 parameters 
(amplitudes and phases of 15 harmonics) are relevant. Also, the WK filter is affected by 
the spectral leakage, which may fail to capture all details of the correlation between 
the witness and the target channels. To solve these problems, we propose to use
the fast Wilson-Daubechies-Meyer transformation (WDM)~\cite{Necula} to split data into the 
frequency sub-bands (time-frequency map). Figure~\ref{WDM} shows an example of the 
WDM time-frequency map for a segment of data from the Hanford detector.
The WDM transformation is orthonormal, invertible 
and has a low spectral leakage, which makes it a unique tool for the regression analysis. 
The data in each frequency sub-band (WDM layer) are a time series used for the construction 
of a single WK filter as described in the beginning of this section. Therefore, instead 
of one long WK filter, 
a bank of much shorter WK filters is constructed. Each filter is trained individually to 
capture details of the target-witness correlation in the corresponding sub-band with just a few 
filter coefficients. For example, a typical analysis in the frequency band 0-2048 Hz involves 
the construction of 2048 independent WK filters with 11 coefficients each. 
Therefore, instead of the inversion of a very large matrix, one has to 
invert 2048 much smaller matrices, which significantly reduces the complexity of 
the regression problem. By applying the inverse WDM transform, clean data $h-x$ in
the time domain can be obtained.

\subsection{Multiple witness channels \label{mwc}}

The GW channel gets contaminated by the environmental noise entering from different physical locations and directions. It is a commonly encountered scenario when a particular 
environmental noise coupled to the target channel is measured by the witness channels situated at different locations. In general, 
when making a prediction over a period of time, data from multiple witness channels are 
expected to provide more complete information about the environment and improve
the estimation of the filter coefficients.
For example, the seismometers are installed at various key locations in the groups of three throughout the LIGO sites, 
measuring seismic noise along specified x, y and z directions. Therefore, a prediction of the seismic contribution to the target 
channel would require a simultaneous use of all three seismic channels.  
The regression analysis addresses the multiple witness channel case by
 extending Equation~\ref{wkhchis}:
\begin{eqnarray} 
&&\chi^2=\sum_{i=L}^{N+L}\Bigg(h[i]-\sum_{j=-L}^{L}a_{j}x_{i+j}-\sum_{k=-L}^{L}b_{k}y_{i+k} \nonumber\\
&&\hspace{2.5cm}-\sum_{m=-L}^{L}c_{m}z_{i+m}+\cdot\cdot\cdot\Bigg)^2. \label{multiwkhchis}
\end{eqnarray}
Where $h$ is the target channel, $x$, $y$, $z$ are the witness channels and $\textbf{a}, \textbf{b}, \textbf{c}$ are the filter coefficients. The corresponding 
Wiener-Hopf equation is
\begin{equation} 
Q
\left[ \begin{array}{cccc}
\textbf{a}\\\textbf{b}\\\textbf{c}\\ \cdot
\end{array} \right]
=
\left[ \begin{array}{cccc}
\textbf{p}_{tx}\\\textbf{p}_{ty}\\\textbf{p}_{tz}\\ \cdot 
\end{array} \right],
\textbf{ } Q\equiv
\left[ \begin{array}{cccc}
R_{xx} & C_{xy} & C_{xz} &\cdot\\ C_{yx} & R_{yy} & C_{yz} &\cdot\\ C_{zx} & C_{zy} & R_{zz} & \cdot\\ \cdot & \cdot & \cdot&\cdot 
\end{array} \right]. \label{multiwkh} 
\end{equation} 
where $R_{xx}/C_{xy}$ are Hermitian Toeplitz matrices of the witness auto/cross correlation
and $\textbf{p}_{hx}$, $\textbf{p}_{hy}$, $\textbf{p}_{hz}$ are the cross-correlation 
vectors between the witness and the target channels. 
The output of the multi-channel filter is given by:
\begin{equation}x[i]=\sum_{j=-L}^{L}a_{j}x_{i+j}+\sum_{k=-L}^{L}b_{k}y_{i+k}+\sum_{m=-L}^{L}c_{m}x_{i+m}.\cdot\cdot\end{equation}

\begin{figure}[htbp]
 \begin{center} 
 \includegraphics[width=0.5\textwidth]{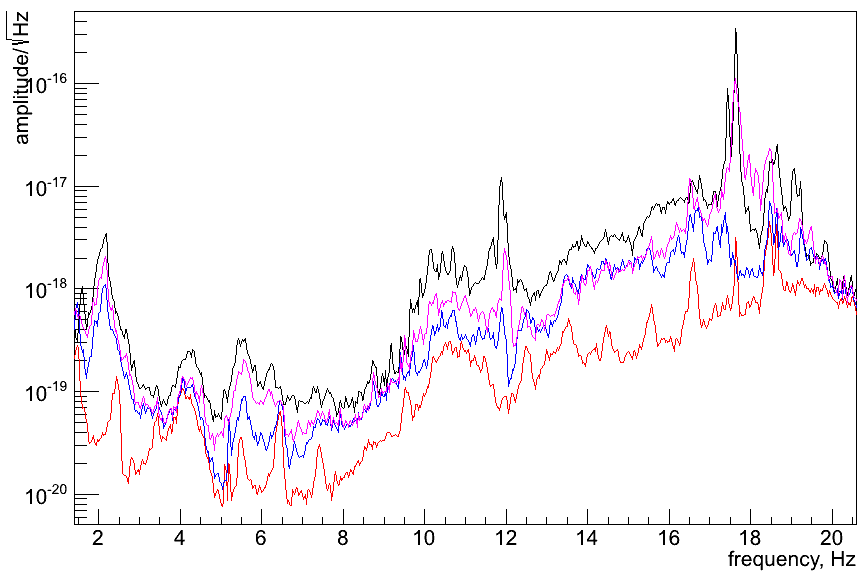}
 \end{center}
\caption{\label{imp_seismic_example}Prediction of seismic noise using 16 coil current channel. Black curve is the original power spectrum. Red curve is 
obtained by conditioning the data using multichannel analysis. Blue curve is obtained by performing single 
channel regression analysis, on the target data, using the CC channels consecutively. Purple curve is a single channel analysis with a different order of the CC channels.}
\end{figure}

The example of improvement in prediction due to the multi-channel regression
is presented in 
figure~\ref{imp_seismic_example}. It shows the regression of seismic noise using the ``coil current'' (CC) channels. The CC channels measure the current in the electromagnets driving 
small magnets attached to the test masses. The main function of these electromagnets is to counteract the seismic motion and hold the test masses in place. Since seismic noise 
at the sites usually affects nearby test masses, at a given time, only a small subset out of total 16 CC channels (positioned at the initial and end test mass) exhibits correlation with 
the GW channel. Therefore, the regression with a single CC channel is not effective. 
One can construct 16 regression cases and apply them consecutively, one CC channel 
at a time. However, as figure~\ref{imp_seismic_example} shows, the best 
performance is obtained with the multi-channel regression when all 16 channels
are used to construct a single regression filter. 


The multi-channel analysis \ref{multiwkhchis} presents two shortcomings. First, if the witness 
channels are highly correlated, the matrix $Q$ can be rank deficient. 
Second, at any given moment a significant fraction of witness channels may not have any measurable 
correlation with the target channel and just add noise at the filter output.
This is demonstrated in table~\ref{noiseincrease}, where we consider 16 witness channels
with random Gaussian noise. In this case the output of the filter is also Gaussian 
noise with the RMS increasing as $\sqrt{M}$, where $M$ is the number of the noisy witness 
channels used in the regression. This case shows, that for a significant number of 
noisy witness channels
the regression analysis may result in the over-fitting of the target data.  
 

\begin{table}
\centering\begin{tabular}{|c|c|c|c|c|c|}\hline
Number of channels &  1  &  2   &  4  &  8  &  16 \\
\hline
RMS of prediction  & .07  & 0.09 & 0.13   & 0.19   & 0.26  \\
\hline
\end{tabular}
\caption{In the absence of correlation between the target and the witness channels 
the RMS of the prediction increases with the number of witness channels due 
to over-fitting, i.e. the analysis makes prediction even in the absence of correlation.}
\label{noiseincrease}
\end{table}

\subsection{Regulators}

\begin{figure}[htbp]
 \begin{center} 
 \includegraphics[width=0.5\textwidth]{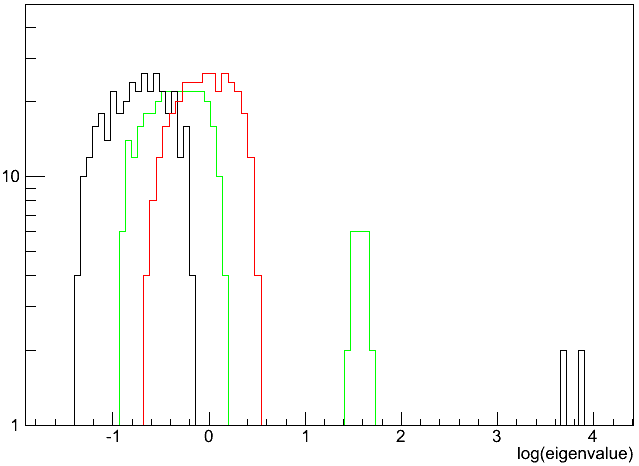}
 \end{center}
\caption{\label{all_eigen_dist}Eigenvalue distribution of matrix $Q$ for some cases. Black - all 16  witness channels carry the same harmonic noise along with 
additional white noise. Red - all the 16 witness channels carry white noise, none of the witness channels are correlated with each other. Green - all the 16 
witness channels carry white noise and all are correlated with each other. }
\end{figure}

Both shortcomings can be addressed by using regulators. 
The equation \ref{multiwkh} can be rewritten as
\begin{equation} 
\left[ \begin{array}{cccc}
\textbf{a}\\\textbf{b}\\\textbf{c}\\ \cdot
\end{array} \right]
=
O\Lambda^{-1}O\left[ \begin{array}{cccc}
c_{hx}\\c_{hy}\\c_{hz}\\ \cdot 
\end{array} \right],
\textbf{ } \Lambda\equiv
\left[ \begin{array}{cccc}
\lambda_0 & 0 & 0 &\cdot\\ 0 & \lambda_1 & 0 &\cdot\\ 0 & 0 & \lambda_2 & \cdot\\ \cdot & \cdot & \cdot&\cdot 
\end{array} \right], \label{rearrangedmultiwkh} 
\end{equation} 
where $O$ is the orthogonal matrix that diagonalizes $Q$, and $\lambda_i$ ($i=0,1,2...$) 
are the eigenvalues arranged in the decreasing order. 
The eigenvalue distribution captures the principal regression components
and identifies a strong correlation in the witness data.  
Figure~\ref{all_eigen_dist} plots the eigenvalue distributions for different regression cases.
Because the witness data are whitened before performing the analysis, the mean of 
the eigenvalue distribution is always equal to one.
When $\lambda_i >> 1$, this is an indication of strong narrow-band noise artifacts
in the witness data. Respectively, when $\lambda_i << 1$, the witness data are
highly correlated and the matrix $Q$ has a lower rank. The small eigenvalues should 
be regulated to avoid un-physical solutions of the WH equation.
The regulated solution is obtained by modifying the matrix $\Lambda$,   
\begin{equation} 
\left[ \begin{array}{cccccc}
\textbf{a}\\\textbf{b}\\\textbf{c}\\ \cdot \\ \cdot \\ \cdot
\end{array} \right]
=
O\Lambda^{-1}O\left[ \begin{array}{cccccc}
c_{hx}\\c_{hy}\\c_{hz}\\ \cdot \\ \cdot \\ \cdot
\end{array} \right],
\textbf{ } \Lambda\equiv
\left[ \begin{array}{cccccc}
\lambda_0&0&0&0&0&\cdot \\ 0&\cdot&0&0&0&\cdot \\ 0&0&\lambda_p&0&0&\cdot\\ \hline0&0&0&C&0&\cdot\\ 0&0&0&0&\cdot&\cdot \\ \cdot&\cdot&\cdot&\cdot&\cdot&C
\end{array} \right]
\lambda_t 
\end{equation}
where the eigenvalues below the threshold $\lambda_t$ a set to some constant $C$.
We distinguish three regulation schemes: \emph{hard} ($1/C=0)$, \emph{mild} ($C=\lambda_0$)
and \emph{soft} ($C=\lambda_p$). 
The proposed regulators handle the problem of the rank deficiency in the matrix $Q$ 
by suppressing the eigenvectors with small eigenvalues thereby preventing un-physical solutions. 
In the case of no correlation between the witness and the target channels, the regulators 
also constrain the system to reduce the RMS of the random noise in the output of the filter. 

\subsection{Effect of regulators on prediction}

The regression parameters and the type of the regulator to use depend on the type of problem on hand. The coupling of the environmental noise to the target and the witness channels usually 
changes with time. 
Hence, no general prescription can be made on the application of the regulation schemes. Below, we discuss figures of merit for some frequently 
encountered cases. In all discussed cases, the target and the witness channels differ by an additional Gaussian noise in the witness channel.

\begin{figure}[htbp]
 \begin{center} 
 \includegraphics[width=0.5\textwidth]{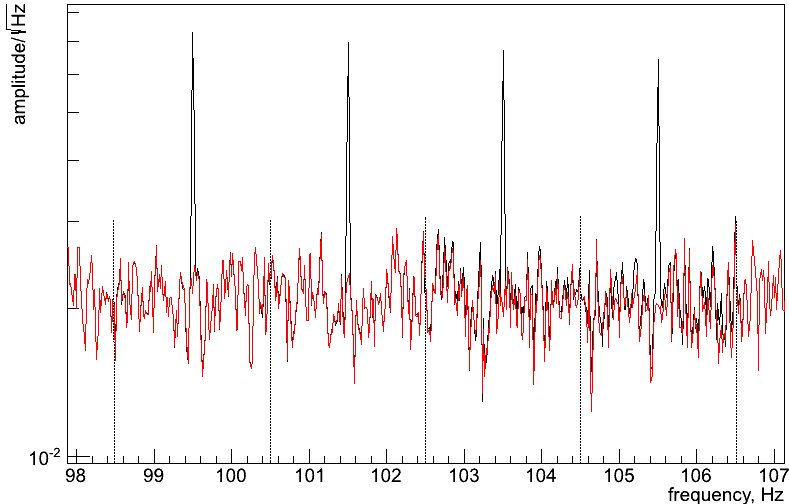}
 \end{center}
\caption{\label{sin_reg_comp}Prediction of simulated monochromatic noise using different regulators with $\lambda_t=1$. Black curve is the original power spectrum. Red 
curve is obtained after conditioning the data. The 99.5~Hz line has been cleaned using 
the hard regulator, the 101.5~Hz line has been cleaned using the mild regulator, 
the 103.5~Hz line has been cleaned using the soft regulator, and the 105.5~Hz 
line has been cleaned with no regulators. }
\end{figure}

1) All three regulators are very close in performance for the prediction of quasi-monochromatic noise. Figure~\ref{sin_reg_comp} compares the regression of simulated monochromatic noise with 
different regulators. As only few eigenvalues are significant in this case, all regulation schemes remove the noise peak. The effect of the over-tuning is clearly 
visible around the peak when no regulator is used. The hard regulator is the best choice for such cases.

2) In the case when all witness channels measure independent broadband noise, the eigenvalue distribution is Gaussian. If none of the witness channels is correlated with 
the target channel, the goal is to reduce the RMS value of the prediction. Table~\ref{white_noise_regulators} shows the performance of various regulators. In this case using 
a hard regulator with a high threshold works the best. On the contrary, if the correlation exists only between one witness channel and the target channel at a time, regulating a 
large number of eigenvectors results in reduced performance, as shown in table~\ref{corr_white_case_2}. A soft regulator or no regulator works best for this case. 
This case is only relevant when the correlation between the witness and the target channel is strong. If the correlation is weak, due to the contamination by uncorrelated witness 
channels, regression analysis does not perform well; for instance, with 16 witness channels and one of them correlated with the target channel with $r_c = .6$, the RMS of the 
residual is 0.8.

3) In the case when all the witness channels have Gaussian noise present and more than one are correlated with the target channel, there is splitting in the eigenvalue 
distribution (figure~\ref{all_eigen_dist}). All the regulators work equally well in this case. For instance, when 8 out of 16 witness channels are correlated with the target channel 
with $r_c=.6$, the RMS of residual is close to .4 for all the regulators and threshold values of up to $\lambda_t = 1.5$.

\begin{table}
\centering
\begin{tabular}{|c|c|c|c|c|c|c|}\hline
Threshold &  0.25  &  0.5   &  0.75  &  1.0  &  1.25  & 1.5 \\
\hline
Hard      & 0.26  & 0.26 & 0.23   & 0.18   & 0.12  & 0.05 \\
\hline
Mild   & 0.26 & 0.26 & 0.24   & 0.20   & 0.18  & 0.17 \\
\hline
Soft   & 0.26 & 0.26 & 0.24   & 0.24   & 0.18  & 0.18 \\
\hline
\end{tabular}
\caption{\label{white_noise_regulators} Effect of regulators when all the 16 witness channel carry Gaussian noise and none of them are correlated with the target channel. 
Table shows the RMS value of the prediction.}
\end{table}

\begin{table}
\centering
\begin{tabular}{|c|c|c|c|c|c|c|}\hline
Threshold &  0.25  &  0.5   &  0.75  &  1.0  &  1.25  & 1.5 \\
\hline
Hard      &  0.152 & 0.152 & 0.381 & 0.617 & 0.811 & 0.953 \\
\hline
Mild      &  0.152 & 0.152 & 0.262 & 0.352 & 0.396 & 0.410 \\
\hline
Soft      &  0.152 & 0.152 &  0.161 & 0.210 & 0.283 & 0.362 \\
\hline
\end{tabular}
\caption{\label{corr_white_case_2} Effect of regulators when only one witness channel out of 16 is correlated with the target channel with $r_c =0.99$. All channels contain 
Gaussian noise. Table shows RMS of the residual.}
\end{table}







\begin{figure}[htbp]
 \begin{center} 
 \includegraphics[width=0.5\textwidth]{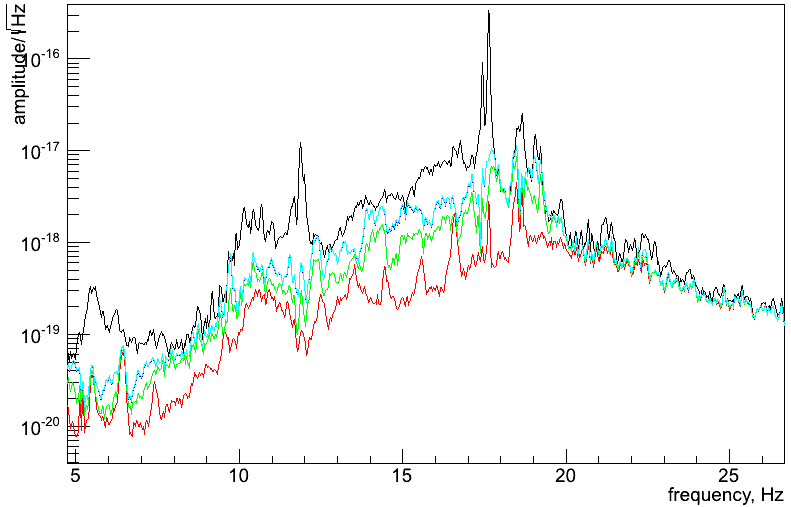}
 \end{center}
\caption{\label{seis_reg_comp} Performance of various regulators (with threshold 0.5) in subtraction of seismic noise using 16 CC channels. Black curve is the original power spectrum. Colored 
curves are obtained after conditioning the data. Red - No regulator, 
Green- Soft regulator, Blue- Mild regulator, and Cyan - Hard regulator. Effect of regulator is clearly visible. It can be deduced that only one CC channel is correlated with the 
GW channel at a time.}
\end{figure}

\section{Applications of WDM regression \label{app}}

There is no one single prescription for the WDM regression. Each regression case depends on a set of available witness channels and type of the noise to be 
predicted. Primary consideration is that witness channels should not be sensitive GW. This consideration can be lifted, when cleaning upconverted noise, if the frequency 
domain of synthetic channels (discussed later) does not overlap with the frequency domain of the channels used in their construction. Below, we consider several typical regression cases performed on the 
data obtained from the Hanford detector.  

\subsection{Regression of noise with linear coupling.}

\begin{figure}[htbp]
 \begin{center} 
 \includegraphics[width=0.5\textwidth]{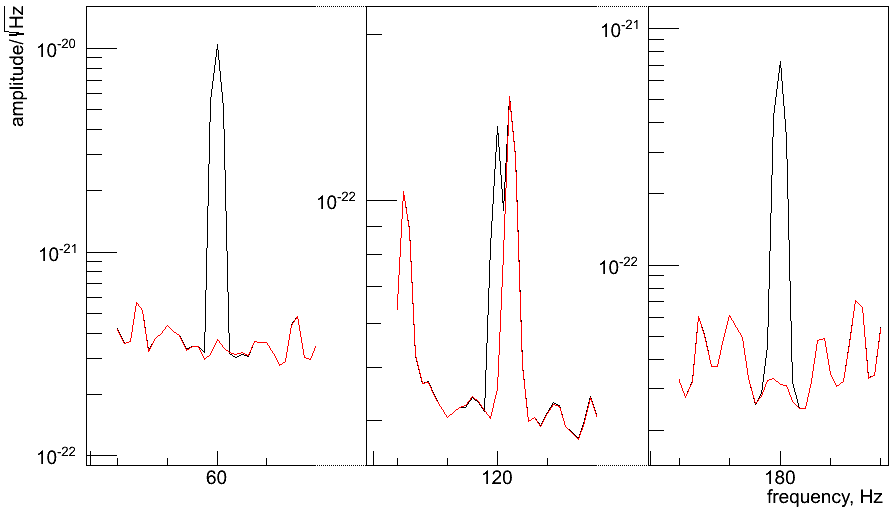}
 \end{center}
\caption{\label{powerlines} Cleaning of power-lines using voltage monitor (H0:PEM-LVEA2\_V1) as the witness channels. Black curve is the original 
spectrum (H1:LDAS-STRAIN); red is obtained after subtracting the prediction.}
\end{figure}

One of the most visible noise artifacts in LIGO detectors are power harmonics 
showing up in the noise spectrum as the quasi-monochromatic lines at multiples of 60 Hz. 
They can be  monitored with the voltage monitors and magnetometers. One or  several 
voltage monitors can be used as witness channels. Figure~\ref{powerlines} shows the noise 
spectrum before and after the application of the WDM regression analysis. 
Note, that the line adjacent to 120 Hz in figure~\ref{powerlines} (middle) is not removed
because it is not monitored by the voltage channels. 

\begin{figure}[htbp]
 \begin{center} 
 \includegraphics[width=0.5\textwidth]{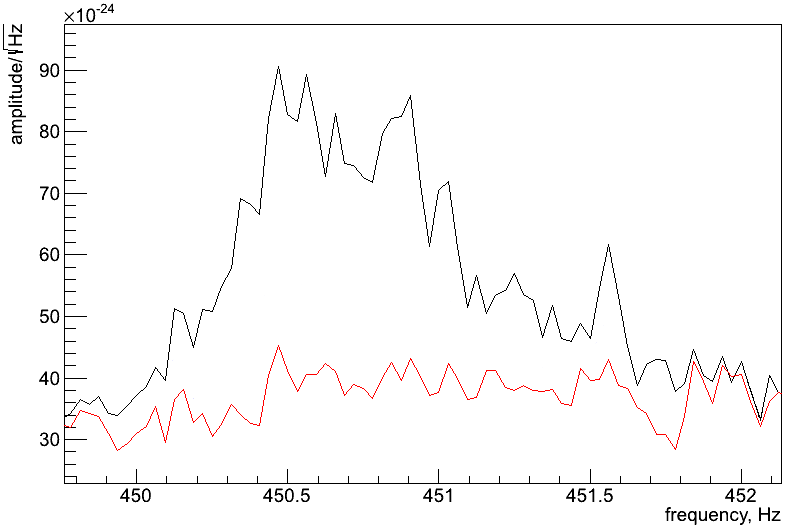}
 \end{center}
\caption{\label{corr_white_case_1_example} Subtraction of mechanical resonances using 14 accelerometers and 2 microphones. Black curve is the original spectrum
(H1:LDAS-STRAIN); red curve is obtained after subtracting prediction using a mild regulator (with threshold 1). 
}
\end{figure}

Other frequently visible features in the LIGO noise spectrum are due 
to the mechanical resonances of the seismic stacks.  These lines are much wider
than the power lines, but they can still be efficiently removed by using regulated
regression with multiple witness channels  (Figure~\ref{corr_white_case_1_example})

A more difficult linear regression case is the prediction of the broadband noise,
like seismic noise. 
Figure~\ref{imp_seismic_example} shows the example of the seismic noise regression 
using 16 coil CC channels (H1:SUS-ITMX\_COIL\_LL, H1:SUS-ITMX\_COIL\_LR, H1:SUS-ITMX\_COIL\_UL, 
H1:SUS-ITMX\_COIL\_UR, H1:SUS-ETMX\_COIL\_LL, H1:SUS-ETMX\_COIL\_LR, H1:SUS-ETMX\_COIL\_UL, H1:SUS-ETMX\_COIL\_UR, H1:SUS-ITMX\_COIL\_LL, H1:SUS-ITMX\_COIL\_LR, H1:SUS-ITMX\_COIL\_UL, 
H1:SUS-ITMX\_COIL\_UR, H1:SUS-ITMX\_COIL\_LL, H1:SUS-ITMX\_COIL\_LR, H1:SUS-ITMX\_COIL\_UL, H1:SUS-ITMX\_COIL\_UR, see Section~\ref{mwc}). We do not
observe a significant rank deficiency of the witness matrix indicating that there is no 
significant correlation between the CC channels. As expected, in this case, the 
best regression is obtained when no regulator is used.

\subsection{Regression of non-linear noise}

\begin{figure}[htbp]
 \begin{center} 
 \includegraphics[width=0.5\textwidth]{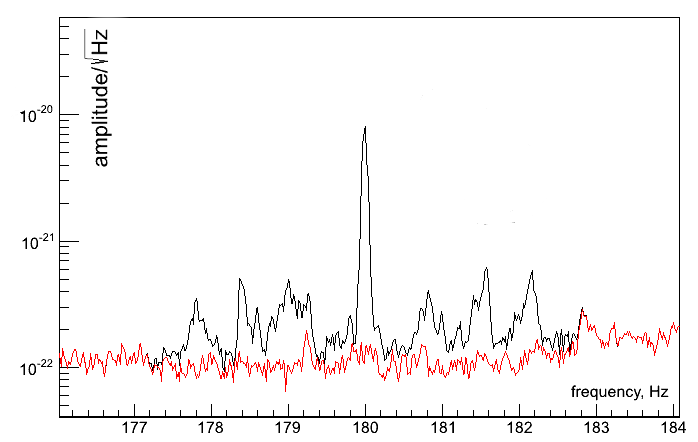}
 \end{center}
\caption{\label{powerlinesbi} Removal of the up-conversion noise around 180~Hz 
power line with 16 synthetic witness channels constructed from 16 coil 
current channels and the voltage monitor \ref{powerlines}. Black curve is the original power spectrum. Red 
curve is obtained after conditioning the data.}
\end{figure}

Some  noise artifacts in LIGO are produced by the interference of two 
(bi-linear noise) or more noise sources in the detector.
The majority of the bi-linear noise cases are due to the up-conversion of 
the low frequency seismic noise, which is observed as the side-bands around 
the power lines, calibration lines, violin modes and other narrow-band features
in the noise spectrum. There are no direct witness channels in LIGO to monitor
such bi-linear noise. As the CC channels provide a good measure of 
the low frequency seismic noise, they can be used along with the other 
channels to synthesize the bi-linear witness channels. 
Such synthetic witness channels are obtained by multiplying the CC time series 
by the carrier time series representing a narrow-band spectral artifact.
For example, when the carrier is one of the power lines, the bi-linear witness
channel can be obtained as a product of the time series from a CC channel and
a voltage monitor or magnetometer. Figure~\ref{powerlinesbi} shows the example 
of the up-conversion noise regression around 180~Hz power line with 
16 synthetic witness channels constructed from the CC channels and one voltage
monitor channel. Usually, such cases need to be regulated to
avoid the regression artifacts.  

\begin{figure}[htbp]
 \begin{center} 
 \includegraphics[width=0.5\textwidth]{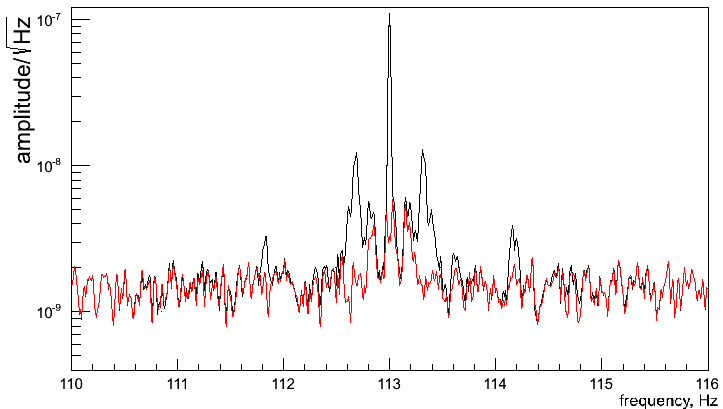}
 \end{center}
\caption{\label{calibi}Removal of up-conversion around 113 Hz calibration line using coil current channels and a simulated noise. Black curve is the original power spectrum. Red 
curve is obtained after conditioning the data.}
\end{figure}

\begin{figure}[htbp]
 \begin{center} 
 \includegraphics[width=0.5\textwidth]{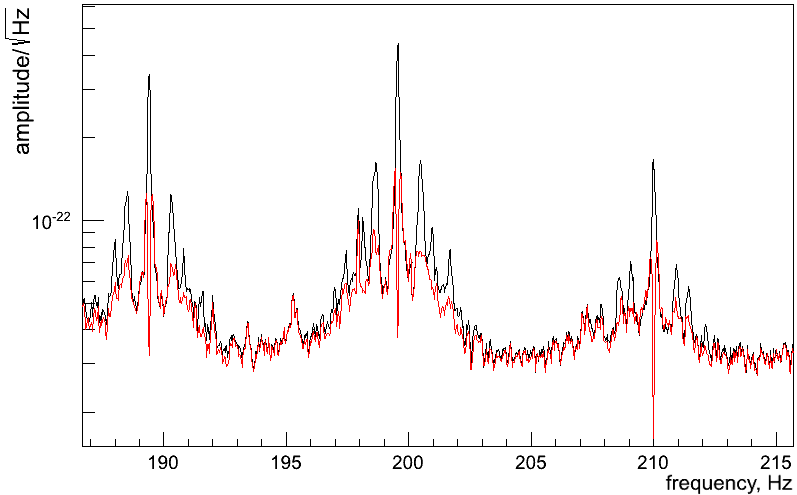}
 \end{center}
\caption{\label{jitterbi}Removal of up-conversion around jitter peaks using coil current channels and self-predicted carrier. Black curve is the original power spectrum. Red 
curve is obtained after conditioning the data.}
\end{figure}

Often, the carrier time series are not readily measured by any PEM channels. 
If the carrier frequency is quite stable, the carrier witness channel can be
simulated with the monochromatic signal. 
Figure~\ref{calibi} shows how this method works for the regression of 
the up-conversion noise around the calibration lines. Also, the carrier witness
channel can be extracted from the strain (target) data with the linear
prediction error (LPE) filter. In this case, the LPE filter is trained 
on the WDM data in the narrow-band ($\sim 1$~Hz) around the carrier frequency. 
The output of the LPE filter is used as the carrier witness channel.
Figure~\ref{jitterbi} shows the subtraction of the seismic up-conversion noise 
around the beam jitter peaks, which were not monitored by any witness channels
during the LIGO science runs. 


\subsection{Monitoring of Gravitational Wave Data}

\begin{figure}[htbp]
 \begin{center} 
 \includegraphics[width=0.5\textwidth]{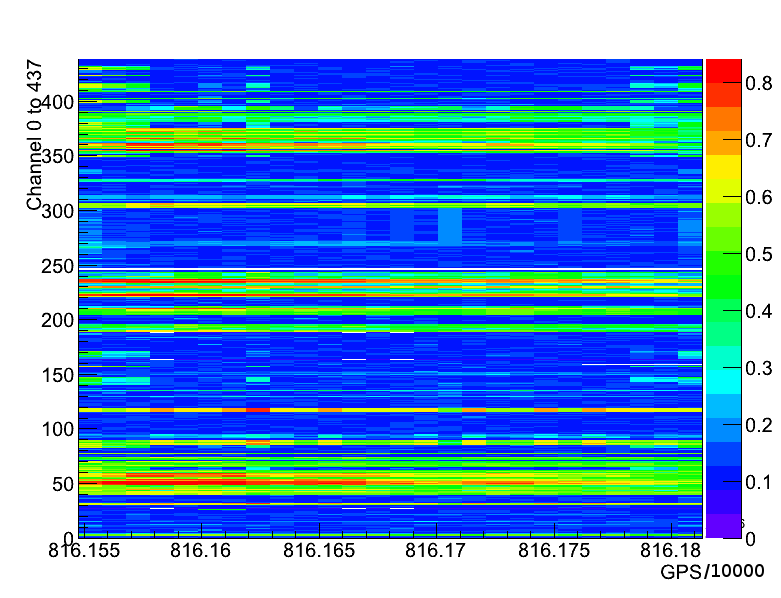}
 \end{center}
\caption{The maximum RMS value as a function of the channel identification number
and GPS time. }
\label{monitoring}
\end{figure}

LIGO has hundreds of auxiliary channels. Some of these channels (like voltage monitors) are very effective in the monitoring of the noise artifacts present in the target data. However, many
channels do not exhibit any significant correlation with the strain channel or this correlation varies significantly over the time. 
The proposed regression analysis can be used to monitor the correlation between the target and the witness channels during the data taking runs. The measure of the correlation is the RMS value 
of the whitened prediction signal, which should be between 0 (no correlation)  and 1 (strong correlation). Since it is hard to visualize all RMS values from hundreds of LIGO channels, 
the maximum RMS value among all the WDM layers in each witness channel can be used to identify  the strongest correlation with the strain channel. Figure~\ref{monitoring} shows the maximum RMS 
value for approximately four hundred LHO channels (all the channels which have sampling rate of 2048 Hz or higher).  To obtain this figure of merit, for each witness channel the 2048 regression 
filters were constructed covering a 1~Hz frequency sub-band each. The RMS values of the filter output were calculated and the maximum value is displayed as a function of the channel identification 
number and the GPS time of the test data segment. In this plot, the contribution of the power lines has been excluded, as they are correlated with many channels. There is no significant 
correlation for most of the tested channels: green and blue areas in the plot where RMS$\leq$0.5. Some auxiliary channels have a strong correlation with the strain channel: 
yellow and red areas in the plot.

There are other possible figures of merit that can be used. For example, the average RMS value of the predicted noise from several WDM layers is a good indicator of the broad-band correlation with 
the strain data. Also, the witness channels can be monitored in groups by using the multi-channel regression. This monitoring method may help to identify channels, which are weakly correlated with the 
GW channel individually, but may reveal a strong correlation as a group.

\section{Conclusion \label{conclusion}}

We present a novel method for regression of LIGO data. Using the LIGO
PEM data estimates the contribution of the environmental noise
into the GW channel.
The regression analysis is performed in the time-frequency domain obtained
with the WDM transform. Banks of the Wiener-Kolmogorov filters are constructed
to capture the details of the witness-target correlation.   
The regression analysis is extended to multiple channel case as, in general, a single witness channel 
may not have sufficient information about the correlation between the GW channel.
 Inclusion of multiple channels is shown to increase the efficiency of the regression analysis, and also present the problems of over-fitting and rank deficiency 
of linear equations. Regulators are introduced to mitigate these problem, thereby, removing the need for the micromanagement of multiple channels used in the analysis. 
Linear noise is regressed by directly accepting the data from the PEM channels as the witness channels, while bi-linear noise is handled by constructing witness channels 
from two or more PEM and/or simulated channels. Regression in the WDM domain along with the freedom to adjust the regulators gives the analysis the flexibility in handling 
variety of noises appearing at various frequencies. Regression analysis can also be used for monitoring data from future LIGO detectors. With hundreds of PEM expected 
to be functional for future detectors, channel monitoring will identify the PEMs having the greatest effects on a detector's performance.


\section{Acknowledgement}

We are thankful to the LIGO group at University of Florida for the lively discussion and creative feedback. We also thank the National Science Foundation for support 
under grant PHY 1205512. This document has been assigned LIGO Laboratory document number P1400004.


\end{document}